\documentclass[reprint,amsmath,amssymb,superscriptaddress,aps,prl,bibnotes]{revtex4-1}
\usepackage{graphicx}
\usepackage{dcolumn}
\usepackage{bm}
\usepackage{color}
\usepackage[dvipdfmx]{hyperref}
\hypersetup{
 setpagesize = false,
 colorlinks = true,
 citecolor = blue,
 linkcolor = blue,
 urlcolor = blue,
 }
\usepackage[all] {hypcap}
\usepackage{comment}
\usepackage{nccmath}

\begin{document}

\title{Soft X-ray Vortex Beam detected by In-line Holography}

\author{Yuta Ishii}
\email{yuta.ishii.c2@tohoku.ac.jp}
\affiliation{Department of Physics, Tohoku University, Sendai 980-8578, Japan}
\author{Kohei Yamamoto}
\affiliation{Division of Electronic Structure, Department of Materials Molecular Science,  Institute for Molecular Science, Okazaki 444-8585, Japan}
\author{Yuichi Yokoyama}
\affiliation{Japan Synchrotron Radiation Research Institute (JASRI/SPring-8), Sayo 679-5198, Japan}
\author{Masaichiro Mizumaki}
\affiliation{Japan Synchrotron Radiation Research Institute (JASRI/SPring-8), Sayo 679-5198, Japan}
\author{Hironori Nakao}
\affiliation{Photon Factory, Institute of Materials Structure Science, High Energy Accelerator Research Organization, Tsukuba 305-0801, Japan}
\author{Taka-hisa Arima}
\affiliation{RIKEN Center for Emergent Matter Science (CEMS), Wako 351-0198, Japan}
\author{Yuichi Yamasaki}
\affiliation{Photon Factory, Institute of Materials Structure Science, High Energy Accelerator Research Organization, Tsukuba 305-0801, Japan}
\affiliation{Research and Services Division of Materials Data and Integrated System (MaDIS), National Institute for Materials Science (NIMS), Tsukuba, 305-0047, Japan}
\affiliation{RIKEN Center for Emergent Matter Science (CEMS), Wako 351-0198, Japan}
\affiliation{PRESTO, Japan Science and Technology Agency (JST)}

\begin{abstract}

We demonstrate the 
in-line holography for soft x-ray vortex beam having an orbital angular momentum.
A hologram is recorded as an interference between a Bragg diffraction wave
from a fork grating and a divergence wave generated by a Fresnel zone plate.
The obtained images exhibit fork-shaped interference fringes, 
which confirms the formation of the vortex beam.
By analyzing the interference image, 
we successfully obtained the spiral phase distribution 
with the topological charge $\ell = \pm 1$.
The results demonstrate that 
the in-line holography technique is promising for the characterization of
topological magnets, such as magnetic skyrmions.

\end{abstract}

\maketitle

A vortex wave function $\psi = \exp (i\ell \phi)$,
where $\phi$ is the azimuthal angle in cylindrical coordinates,
is an eigenstate of the orbital angular momentum (OAM) operator
$\hat{L}_z = -i\hbar \partial/\partial \phi$ with an eigenvalue $\ell \hbar$.
Therefore, spiraling wavefront of a particle provides an OAM around the propagation axis, 
and the vortex beam is characterized by the topological charge $\ell$ defined as
\begin{equation}
    \ell = \frac{1}{2\pi} \oint_C \nabla \theta({\bm r}) \cdot d{\bm r},
\end{equation}
where  $\theta$ represents a phase distribution, and $C$ represents a closed loop surrounding the propagation axis.
The revelation of a
vortex beam on visible light generated in free space \cite{allen1992orbital}
has attracted significant interest
in diverse research fields \cite{shen2019optical}, 
including super-resolution microscopy \cite{Xiaodong2018},
optical tweezers \cite{DavidG2003}, and quantum information processes \cite{Gabriel2007}.

Soft x-ray vortices have 
also been investigated for their generation 
from undulators \cite{Bahrdt_2013}, spiral zone plates \cite{sakdinawat2007soft}, and diffractive optics \cite{lee2019laguerre}.
Soft x-rays tuned to the absorption energy of an element
are widely used to investigate the magnetic state of 
materials  containing a 3$d$ or 4$f$ transition element
owing to their high sensitivity 
to the spin density of unoccupied states.
Furthermore,
soft x-ray diffraction with coherent wavefronts is a suitable technique 
for the real-space observation of magnetic textures \cite{yamasaki2015dynamical,ukleev2018coherent,ukleev2019element}.
Hence, soft x-ray vortices are expected to be  a new type of 
imaging probes or manipulating tools 
for topological magnets, such as magnetic skyrmions
\cite{van2007prediction, van2015interaction, fujita2017ultrafast, yang2018photonic}.

Clarifying  the phase distribution of vortex beams
would be crucial for
characterizing and visualizing
topological features, such as phase singularity, on the magnets, 
with the use of soft x-ray vortices.
However,
it is impossible to directly obtain
the phase information of soft x-rays by observing the intensity.
For visible light, 
an interference measurement
can be performed to visualize the phase distribution of vortex beam 
using optical elements such as beam-splitters and mirrors \cite{MHarris1994, Wang_2009}.
Meanwhile, 
the lack of optical elements for soft x-rays, such as a suitable mirror,
renders it difficult to perform an interference experiment.
Another interference method is in-line (Gabor's) holography,
in which a reference wave  and a diffracted wave scattered 
by an object share the same optical axis.
The usefulness of this technique has been demonstrated in electron vortex beams  
for obtaining spiral phase distributions \cite{saitoh2013measuring, Harada_2020}.
Hence, in the present study, we focus on observing of the phase information 
of soft x-ray vortices generated by a diffractive optical element
via coherent soft x-ray in-line holography.

\begin{figure}[t]
\centering
\includegraphics[clip,width=8cm]{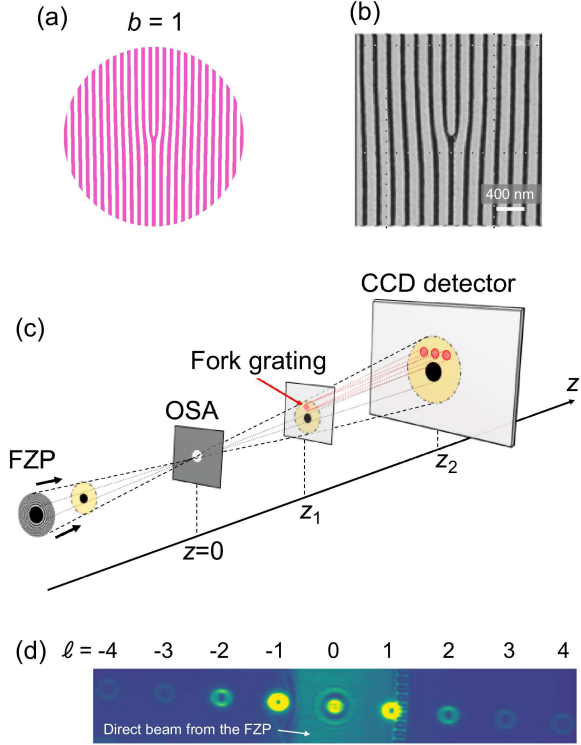}
\caption{(a) Schematic of a fork grating with $b = 1$. 
(b) Scanning electron microscopy image of the central part of the fork grating. 
Ta metal (gray) is deposited on a membrane of Si$_3$N$_4$ used in this study. 
(c) Experimental setup for coherent soft x-ray in-line holography. 
Incident x-rays are focused using a Fresnel zone plate (FZP). 
The first-order diffraction from the FZP is selected by 
order sorting aperture (OSA) placed at the focal point.
The zeroth direct beam was stopped by a center stop on the FZP and the OSA.
The Bragg diffraction waves, 
which occurred at the fork grating located downstream of the focal point, 
interfered with waves transmitted outside the grating. 
(d) Diffraction pattern from fork grating without interference, 
which shows the diffraction up to the fourth order.}
\label{fig:fig1}
\end{figure}

A vortex beam for the Laguerre-Gaussian mode \cite{Beijersbergen1993} with 
the zeroth radial index and an integer topological charge $\ell = nb$ 
can be produced at the $n$-th Bragg diffraction from a fork grating with a topological number $b$ \cite{Bazhenov1990,JArlt1998}.
Phase distribution in the fork grating is expressed as
\begin{equation}
    \varphi (\rho,\phi) = b \phi + \frac{2\pi}{d}\rho\cos \phi
\end{equation}
in the cylindrical coordinate $(\rho,\phi,z)$,
with $d$ being the pitch of the grating far from the center ($\rho =  0$).
The binarized amplitudes corresponding to opaque and transparent regions are
\begin{eqnarray}
    a(\varphi) = 
    \left\{
    \begin{array}{l}
      1, ~~  \text{if}~\text{mod}(\varphi,2\pi) < \pi, \\
      0, ~~ \text{otherwise},
    \end{array}
  \right.
\end{eqnarray}
which results in a pattern  shown in Fig.~\hyperref[fig:fig1]{\ref*{fig:fig1} (a)} for $b = 1$.
The transmitted field $u_{1}$ is calculated in the Fourier series as
\begin{equation}
    u_{1} = u_0 a(\varphi) = u_0\sum_{n=-\infty}^{\infty} A_n e^{-in\frac{\pi}{2}} e^{in\varphi}
\end{equation}
with
\begin{equation}\label{eq:An}
   A_n = \frac{\sin(n\pi/2)}{n\pi}
\end{equation}
and the incident field $u_0 = u_0(\rho,\phi)$  \cite{ZGongjian2018}.
The far field diffraction output $u_2$ is derived from the Fourier transform $\mathcal{F}[u_1]$, 
which results in
\begin{equation}
    u_2(f_x,f_y)\propto   \sum_{n=-\infty}^{\infty}
    u_0 e^{in(b\phi_{kn}+  \frac{\pi}{2}(b-1))}    
    A_n\mathcal{J}_{nb}(\rho_{kn}),
\end{equation}
where $(\rho_{kn},\phi_{kn})$ denotes the cylindrical coordinate with $(f_x,f_y)=(2\pi n/d,0)$ as the origin.
$\mathcal{J}_{nb}(\rho_{kn})$ can be calculated as
\begin{equation}\label{eq:bessel}
    \mathcal{J}_{nb}(\rho_{k}) = \int_0^a \rho J_{nb}(\rho_{k}\rho) d\rho
\end{equation}
with the $n$-th Bessel function of the first kind $J_{nb}(x)$, and the radius of the grating $a$.
The result indicates that the diffracted wave at the $n$-th Bragg diffraction has an OAM with 
$L_z = \ell \hbar = nb\hbar$.

The soft x-ray diffraction measurement does not provide any information about
 the phase factor $e^{i\ell\phi_k}$ 
as only the intensity $I = |u_2|^2$ is observable.
Therefore, we observed the interference 
between diffracted waves and the divergent waves from the Fresnel zone plate (FZP), 
as shown in Fig.~\hyperref[fig:fig1]{\ref*{fig:fig1} (c)}, 
which illustrates the schematics for the in-line holography setup.
Soft x-ray experiments were performed  at 
soft x-ray undulator beamlines 
BL-13A and BL-16A of the Photon Factory, KEK, Japan.
The wave length of soft x-ray was tuned at $\lambda = 1.6$ nm, which ideally
resulted in a focal length of 1.5 mm via the FZP with
 outer and center beam stopper radii of 60 $\mu$m and 30 $\mu$m, respectively.
The higher diffractions from the FZP were filtered by an order sorting aperture (OSA)
with a radius of 5 $\mu$m located at the focal position of the FZP.
A fork grating with $b = 1$
 was made from 
Ta metal with the thickness of 300 nm deposited on the membrane of Si$_3$N$_4$.
Figure \hyperref[fig:fig1]{\ref*{fig:fig1} (b)} shows the scanning electron microscopy image of the grating.
The grating has a nominal 200-nm period and a radius of $a =$ 2.5 $\mu$m.
Not patterning outside the grating
enables the reference waves to pass through
the grating and interfere with the diffracted waves.
The diffraction pattern was recorded using an in-vacuum CCD camera 
(PMI2048,  2048 $\times$ 2048 pixel, pixel size 13 $\mu$m, Teledyne Princeton instrument)
installed at a distance of $z_2$ = 220 mm from the focal position.

\begin{figure}[t]
\centering
\includegraphics[clip,width=8.2cm]{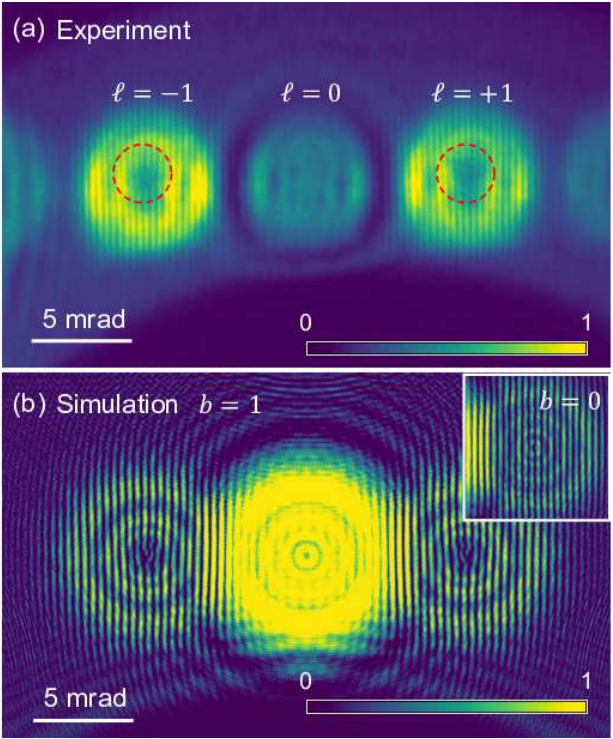}
\caption{(a) Experimental result of the in-line holography for the diffraction of the vortex beam interfering with the transmitted reference beam. 
Dashed circles indicate fork-shaped patterns appearing at Bragg diffractions with
$\ell=-1$ and  $\ell=+1$.
The length of the scale bar represents the angle of 5 mrad  measured from the fork grating.
(b) Simulated interference pattern from a fork grating with $b$ = 1,  and
that from a rectilinear grating ($b$ = 0) is shown in the inset. }
\label{fig:fig2}
\end{figure}

Figure~\hyperref[fig:fig1]{\ref*{fig:fig1} (d)} shows
the diffraction pattern from the fork grating 
placed at the  right part of annular diffraction from the FZP.
In this case, 
the diffraction from the grating is observed without superimposing 
on the transmitted waves.
From the first to fourth Bragg diffraction,
each exhibits a ring-like intensity distribution,
which is characteristic of vortex beams, 
and would have an OAM of $L_z = n \hbar$.
Based on Eq.~\hyperref[eq:An]{(\ref*{eq:An})}, 
the diffraction of $n$ = 2, 4 should vanish,  
provided that the grating forms an ideal square wave. 
The appearance of even-number order Bragg diffraction
 is likely because the widths of the transparent and opaque regions do not match perfectly.

Figure \hyperref[fig:fig2]{\ref*{fig:fig2} (a)}
shows the diffraction pattern when the grating is 
placed at $z_1$ = 680 $\mu$m away from the focal point and on the up-side
of the annular diffraction waves from the FZP.
In this case, interference pattern is observed because
the first diffraction from the grating overlapped with the waves from the FZP.
Intensity modulations along the horizontal orientation
are shown clearly, which can not be observed in the absence of interference.
In addition, upper half of the diffraction pattern 
contains an additional stripe compared with the lower half, 
and then a fork-shaped pattern appears at the center.
Such a pattern implies 
the formation of vortex beams
with the phase rotating from 0 to 2$\pi$ along the azimuthal angle direction.
We simulated an interference pattern 
using the scaled Fast Fourier transform method \cite{David1991}, 
assuming an ideal FZP and a fork grating with $b = 1$.
Figure \hyperref[fig:fig2]{\ref*{fig:fig2} (b)} shows the simulated interference pattern,
which agrees reasonably well with the experimental result.
Furthermore, we simulated the case for a rectilinear grating ($b$ = 0)
as shown in the inset of Fig.~\hyperref[fig:fig2]{\ref*{fig:fig2} (b)},
confirming no fork-shaped pattern at the center.

As similar intensity patterns have been observed in visible light and electron vortex beams 
\cite{MHarris1994, Harada_2020},
it is known that  such interference patterns require a reference beam from an incline direction.
To gain a further understanding 
of the intensity distribution shown in Figs.~\hyperref[fig:fig2]{\ref*{fig:fig2} (a) and \ref*{fig:fig2} (b)},
we performed an analytical calculation
assuming a divergent wave from a point light source at the focal position of the FZP.
It was assumed 
that a point source existed at the origin, and the grating within the $z = z_1$ plane 
as well as the detector 
within the $z = z_2$ plane in the Cartesian coordinates
as shown 
in Fig.~\hyperref[fig:fig1]{\ref*{fig:fig1} (c)}.
The direct light field $u^d_2$ at the detector plane $(x_2, y_2, z_2)$ is expressed as
\begin{eqnarray}
    u^d_2(x_2,y_2,z_2) = \frac{u_0 e^{ikr_2}}{r_2}\sim \frac{u_0}{z_2}e^{ik\left(z_2+\frac{x_2^2+y_2^2}{2z_2}\right)},
\end{eqnarray}
where $r_2$ is the distance between a detection point and the point source;
furthermore we used 
the paraxial approximation $(x_2^2+y_2^2\ll z_2^2)$.
The diffracted wave from the grating $u^g_2(x_2,y_2,z_2)$ is expressed as
\begin{align}
    &u^g_2(x_2,y_2,z_2) = -\frac{i}{\lambda}\iint_G \frac{e^{ikr}}{r}u_1(x_1,y_1)dx_1dy_1,
\end{align}
with $G$ denoting the integral within the grating region, 
and $r = \sqrt{ (x^{\prime}_1 - x_2)^2 + (y^{\prime}_1 - y_2)^2 + (z_1 - z_2)^2}$
denoting the distance between a detection point and a position 
$Q  =(x^{\prime}_1, y^{\prime}_1, z_1)$
on the grating where a x-ray passes through.
The diffraction pattern of these two waves can be calculated as
\begin{eqnarray}\label{eq:intensity}
    I(x_2,y_2) = |u_2^d|^2 + |u_2^g|^2 + u_2^du_2^{g*} + c.c.,
\end{eqnarray}
where $c.c.$ represents the complex conjugate.
The interference term of the $n$-th Bragg diffraction
from the fork grating with $b$ = 1  is calculated as
\begin{align}\label{eq:intn}
I^{\text{inter}}_n = 2C_n|\mathcal{J}^\prime_{n}|  \sin\left\{kR_2+n\phi_{fn} + \alpha_{n} \right\},
\end{align}
with $C_n=2A_n\pi u_0^2 /\{\lambda z_1z_2(z_2-z_1)\}$, 
\begin{align}\label{eq:Jn}
    &    \mathcal{J}^\prime_{n}(\rho_{fn}) = \int_0^a e^{if_0\rho^2} \rho J_{n}(\rho_{fn}\rho) d\rho \equiv  |\mathcal{J}_{n}^\prime (\rho_{fn})|e^{i\alpha_{n}},
\end{align} 
and 
\begin{align}
    R_2=\frac{z_1}{2z_2(z_2-z_1)}\left\{\left(x_2-\frac{z_2}{z_1}x_g\right)^2+\left(y_2-\frac{z_2}{z_1}y_g\right)^2\right\},
\end{align} 
where ($\rho_{fn}$, $\phi_{fn}$) is the local cylindrical coordinate  with 
$(x_{2n},y_{2n})=(z_2x_g/z_1+ 2n\pi(z_2-z_1)/kd, z_2y_g/z_1)$ as the origin, 
which corresponds to the center of the $n$-th Bragg diffraction,
and $(x_g,y_g,z_1)$ is the center of the grating.
The integral in Eq.~\hyperref[eq:Jn]{(\ref*{eq:Jn})}
is different from that in Eq.~\hyperref[eq:bessel]{(\ref*{eq:bessel})}
in terms of the factor of $e^{if_0\rho^{2}}$ with $f_0 = kz_2/2z_1(z_2-z_1)$, 
which originates from the near-field diffraction, 
$i.e.$ the Fresnel diffraction \cite{MBorn}.
Eq.~\hyperref[eq:intn]{(\ref*{eq:intn})} includes the term  of $n\phi_{fn} = \ell \phi_{fn}$ 
in the sine function, 
which imposes the spiral phase information on the observable intensity,
thereby resulting in 
the fork-shaped pattern at the center of the diffraction.
$R_2$ varies in the radial direction from the zero-order diffraction, 
which produces stripe-like intensity modulations along the horizontal direction.
Furthermore, the $\alpha_{n}(\rho_{fn})$ results in modulations in the radial direction
from $(x_{2n},y_{2n})$,
which is independent of the azimuthal angular direction.
These features are observed in both the experimental and simulated results.

\begin{figure}[t]
\centering
\includegraphics[clip,width=9cm]{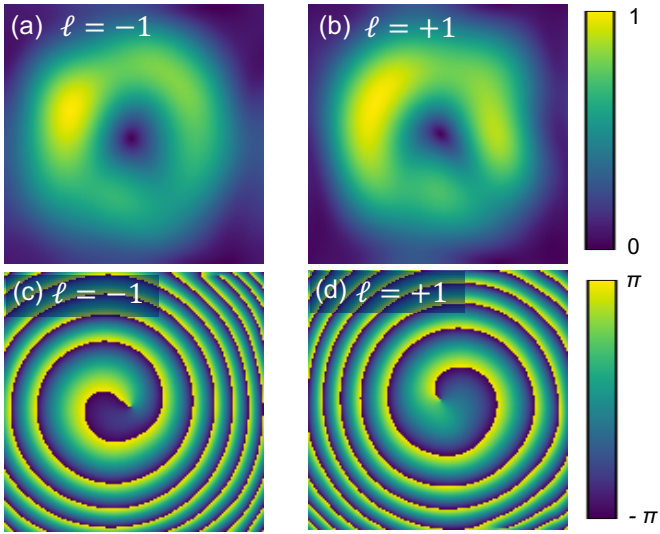}
\caption{(a)(b) Contour of the absolute value of $I^{\prime}_n$ obtained by the way described 
in the text for Bragg diffractions with (a) $\ell = -1$ and (b) $\ell = +1$, 
which are shown in Fig.~\hyperref[fig:fig2]{\ref*{fig:fig2} (a)}.
(c)(d) Spiral phase distribution for (c) $\ell = -1$ and (d) $\ell = +1$ vortex diffraction waves.}
\label{fig:fig3}
\end{figure}

The Fourier transform $\mathcal{F}$ of the $n$-th Bragg diffraction $I_n$ is expressed as
\begin{align}
    \mathcal{F}[I_n (x_2, y_2)] = G_n^0(0,0) + G_n^+(q_x,q_y) +G_n^-(-q_x,-q_y)
\end{align}
where $G_n^0$ and $G_n^\pm$ are derived from non-interference and interference terms,
respectively, 
and $q_x = 2\pi n z_1/dz_2$, $q_y \sim 0$.
The spiral phase distribution for the $\pm n$-th diffraction from the grating with $b = 1$
can be extracted 
by filtering $G_n^\pm(q_x,q_y)$,
followed by the inverse Fourier transform $\mathcal F^{-1}$, 
and multiplying the term $e^{-ikR_2}$, as follows:
\begin{align}
    I^{\prime}_{\pm n}  &= e^{-ikR_2} \mathcal F^{-1}[G_n^\pm] \\
            		 &= -i C_n |\mathcal{J}^\prime_{n}(\rho_{f,\pm n})| e^{i( \pm \ell \phi_{f,\pm n} + \alpha_{\pm n} )},
\end{align}
with the double sign in the same order.
Figures \hyperref[fig:fig3]{\ref*{fig:fig3} (a) and \ref*{fig:fig3} (b)}
present absolute values of $I^{\prime}_n$
for the Bragg diffraction obtained in our experiment,
where the stripe-like intensity modulation vanished for both cases.
The intensity patterns are similar to  
those of the Bragg diffractions for $\ell = \pm1$ 
without interference, as shown in Fig.~\hyperref[fig:fig1]{\ref*{fig:fig1} (d)},
in which a singularity appeared at the center owing to the term of $|\mathcal{J}^\prime_{n}|$.
Furthermore,
we successfully obtained
the spiral phase distribution for $\ell = -1$ and $\ell = +1$ vortex beams 
by extracting the phase term of $I^{\prime}_n$,
as shown in Figs.~\hyperref[fig:fig3]{\ref*{fig:fig3} (c) and \ref*{fig:fig3} (d)}, respectively, 
where the rotating direction is reversed between the $\ell=\pm 1$ diffraction waves.
In addition, 
the phase modulation in the radial direction due to the term of $e^{i\alpha_{\pm n}}$
yielded Fermat's spiral like pattern as observed,
which originates from the use of a divergent light source.

To summarize, 
we successfully  observed the
intensity and phase distributions for soft x-ray vortex beams
generated by fork grating via in-line holography.
Our results are expected to 
pave the way for
topological magnetic materials research using the soft x-ray OAM beam,
including visualization of the topological features
via coherent soft x-ray diffraction imaging.

We thank M. Hatayama and J. Sasakura (NTT-AT) for preparing the fork-shaped grating.
 This work was supported  in part by PRESTO Grant Number JP-MJPR177A,
by Grant-in-Aid for Scientific Research
Nos. JP16H05990, JP17K05130, JP19H04399, JP19K23590, JP20H04458,
MEXT Quantum Leap Flagship Program (MEXT Q-LEAP) Grant Number JPMXS0120184122, and by the Research Foundation for Opto-Science and Technology.
Soft X-ray scattering work was performed under 
 the approval of the Photon Factory Program Advisory Committee (Proposal No. 2015S2-007, 2018S2-006, 2019G590, 2019PF-22).


\bibliographystyle{apsrev4-1}
\bibliography{apssamp}

\end{document}